\documentclass[aps,prd,,twocolumn,superscriptaddress]{revtex4-1}

\usepackage{graphicx}
\usepackage{amsmath}

\begin{document}

\title{Resonator power to frequency conversion in a cryogenic sapphire oscillator}
\vfill
\author{Nitin R.~Nand}
\affiliation{ARC Centre of Excellence for Engineered Quantum Systems, School of Physics, The University of Western Australia, 35 Stirling Highway, Crawley 6009, Western Australia, Australia\\}
\author{Stephen R. Parker}
\email{stephen.parker@uwa.edu.au}
\affiliation{School of Physics, The University of Western Australia, 35 Stirling Highway, Crawley 6009, Western Australia, Australia\\}
\author{Eugene N.~Ivanov}
\author{Jean-Michel~le~Floch}
\affiliation{ARC Centre of Excellence for Engineered Quantum Systems, School of Physics, The University of Western Australia, 35 Stirling Highway, Crawley 6009, Western Australia, Australia\\}
\author{John G.~Hartnett}
\affiliation{School of Physics, The University of Western Australia, 35 Stirling Highway, Crawley 6009, Western Australia, Australia\\}
\affiliation{Institute of Photonics and Advanced Sensing, School of Chemistry and Physics, University of Adelaide, North Terrace, Adelaide 5005, South Australia, Australia\\}
\author{Michael E. Tobar}
\affiliation{ARC Centre of Excellence for Engineered Quantum Systems, School of Physics, The University of Western Australia, 35 Stirling Highway, Crawley 6009, Western Australia, Australia\\}

\date{\today}

\begin{abstract}
We report on the measurement and characterization of power to frequency conversion in the resonant mode of a cryogenic sapphire loaded cavity resonator, which is used as the frequency discriminating element of a loop oscillator circuit. Fluctuations of power incident on the resonator leads to changes in radiation pressure and temperature in the sapphire dielectric, both of which contribute to a shift in the resonance frequency. We measure a modulation and temperature independent radiation pressure induced power to frequency sensitivity of -0.15 Hz/mW and find that this is the primary factor limiting the stability of the resonator frequency.
\end{abstract}

\maketitle

A Cryogenic Sapphire Oscillator (CSO)~\cite{hartnett2006,locke2008,hartnitin32012} uses a Sapphire Loaded Cavity (SLC) resonator as the frequency discriminating element of a Pound lock loop~\cite{pound1946} to generate a microwave frequency output signal. The fractional frequency stability of a CSO output at 11.2~GHz is state-of-the-art for integration times up to $10^{3}$ seconds and is rivaled only by microwave signals generated through optical-comb division~\cite{fortier2011} for short integration times (up to 5~seconds). It has been speculated~\cite{locke2008} that the intrinsic noise of the oscillator Pound lock electronics limits the CSO performance over short integration times (up to 10 seconds) while fluctuations of the power incident on the resonator limits stability for integrations times from 10 seconds onwards, until the longer term effects of frequency drift due to ambient temperature cycles become dominant. Recent work~\cite{hartnitin32012} reported that the frequency stability at 1~second integration time appeared to be independent of the level of power incident on the resonator (from 0.1~mW to 0.8~mW) and decreasing the level of power incident on the resonator lowered the flicker noise floor encountered around integration times of $10^{3}$ seconds. In an effort to understand what currently limits the frequency stability of a CSO, we have investigated how power fluctuations in the cryogenic SLC are converted into frequency fluctuations of the resonant electromagnetic mode.

\begin{figure}
\centering
\includegraphics[scale=0.48,keepaspectratio=true]{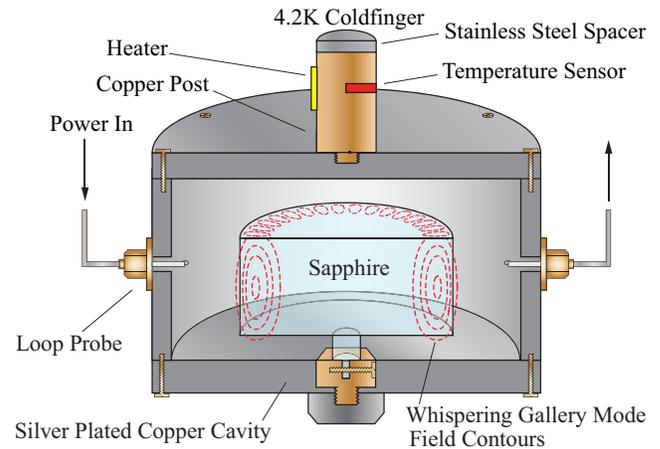}
\caption{(color online) Cross-section of a sapphire loaded silver plated copper cavity used in a cryogenic sapphire oscillator. The resonant mode electromagnetic field patterns are outlined in red.}
\label{fig:SLC}
\end{figure}

The SLC (Fig.~\ref{fig:SLC}) features a HEMEX grade cylindrical sapphire crystal with a diameter of 51.0~mm and a height of 30.0~mm. A spindle protrudes from the central axis of the crystal and is used to clamp the sapphire inside a silver plated copper cavity, which is cooled to cryogenic temperatures in a low vibration pulsed tube cryostat~\cite{wang2010}. A sensor and heater embedded in the copper post are used to control the temperature. Microwave magnetic loop probes excite standing wave patterns, known as Whispering Gallery (WG) modes in the crystal with the majority of the electromagnetic energy confined within an outer ring of the sapphire. This high level of field confinement combined with the low loss tangent of sapphire at cryogenic temperatures results in electrical quality factors of the order 10$^{9}$. Small concentrations of paramagnetic ions in the sapphire create a temperature-dependent magnetic susceptibility in the material~\cite{hartnett1999}. The resonance frequency of the WG mode is dependent upon both the susceptibility and the permittivity of the dielectric crystal. At temperatures below 15~K the magnitude of these dependencies are similar but their signs are opposite, this creates a frequency-temperature turning point where the resonance frequency of the SLC is first order insensitive to temperature variations.

Fluctuations in the power incident on the sapphire create fluctuations of the SLC resonance frequency via two mechanisms. The first is due to a change in the stored energy density, and hence radiation pressure, in the sapphire causing a strain induced shift in the dielectric constant and thus the resonance frequency~\cite{braginsky1985}. The second arises from the change in power heating or cooling the crystal leading to frequency shifts as discussed earlier. The magnitude and sign of the power to frequency conversion efficiency, $\partial f/\partial P$, due to changes in radiation pressure was measured with a previous generation of CSO~\cite{chang1997}, where the resonator was cooled using a liquid helium bath in a large dewar. By altering the power incident on the resonator and monitoring the change in heater power in the resonator temperature control system the change in power dissipated in the resonator can be inferred. The corresponding frequency shift was measured by either probing the resonance with a swept signal generated by an externally referenced synthesizer or by monitoring the beat frequency (on the order of kHz) of the output of two loop oscillators, with one oscillator using the resonator under test as the frequency discriminating element of the Pound lock loop. The experiment was performed using discrete power shifts ranging from less than 1~mW to 20~mW, with the maximum amount of power dissipated in the resonator never exceeding 20~mW. These are relatively large levels of power compared to the typical level of power incident on a CSO resonator ($\approx$~0.1-0.2~mW). The values of $\partial f/\partial P$ obtained are for long time scales where the CSO frequency stability is limited by ambient temperature fluctuations. The current generation of low vibration pulsed tube cryocooler CSOs also utilize a different resonant mode in the SLC and thus have a different frequency sensitivity to radiation pressure changes. In order to measure $\partial f/\partial P$ for the current resonant mode over a range of Fourier frequencies at lower power levels and to determine if the radiation pressure induced frequency shift is both temperature and frequency independent a measurement technique was developed.

\begin{figure}[!t]
\centering
\includegraphics[scale=1.35,keepaspectratio=true]{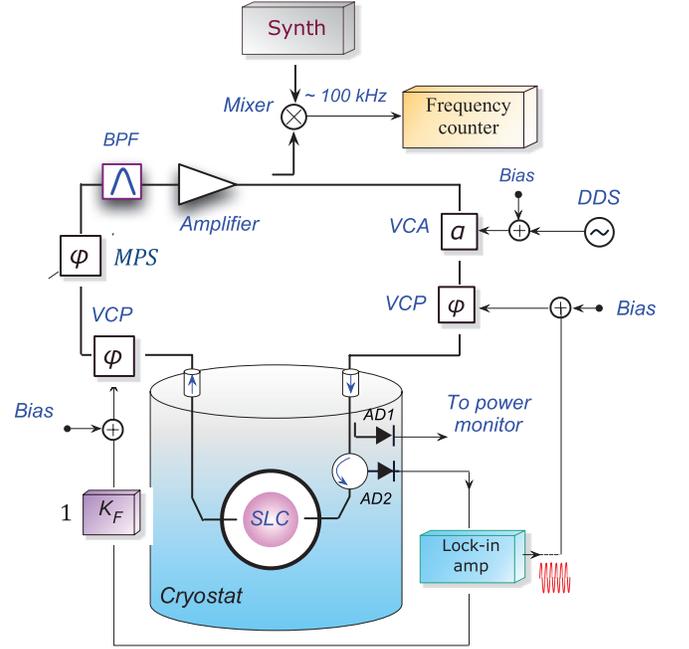}
\caption{(color online) A cryogenic SLC and loop oscillator circuit to measure $\partial f/\partial P$ in the time domain. Coherent sine wave modulation is produced by an arbitrary waveform generator (DDS) and applied to the bias of a voltage controlled attenuator (VCA). A frequency counter logs the beat ($\approx$100~kHz) of the loop oscillator output and a synthesizer referenced to a H-maser. Voltage Controlled Phase shifters (VCP), Mechanical Phase Shifter (MPS), Band Pass Filter (BPF), Amplitude Detector (AD2), lock-in amplifier and filter (K$_{\text{F}}$) are part of the oscillator frequency control system.}
\label{fig3}
\end{figure}

It is possible to calculate the expected radiation pressure induced frequency shift for a specific resonant mode of the SLC, which for the current iteration of the CSO is a WGH$_{16,0,0}$ mode where over 96$\%$ of the electromagnetic energy is contained in the dominant electric axial field. The energy, or power, of the field is distributed around the effective mode volume ($V$) and exerts pressure parallel to the crystal axis ($z$) causing it to strain, such that
\begin{equation}
\partial z / \partial P = (z Q) / (2\pi f VY),
\label{eq:strain}
\end{equation}
where $Y$ is the appropriate Young's modulus. Strain in the sapphire dielectric changes the permittivity and hence the resonance frequency. This is an anisotropic effect and for an axial strain the resulting frequency shift is given by~\cite{locke2004}
\begin{equation}
\partial f / \partial z =\frac{f}{z}\left(\frac{\nu}{2}P_{\epsilon_{r}}K_{\epsilon_{r}}-\frac{1}{2}P_{\epsilon_{z}}K_{\epsilon_{z}}+\nu P_{r}-P_{z}\right),
\label{eq:dfdz}
\end{equation}
where $\nu$ is Poisson's ratio for sapphire (0.3), $P_{\epsilon_{r}}$ and $P_{\epsilon_{z}}$ are the radial and axial electric filling factors, $K_{\epsilon}$ is the fractional change in dielectric constant per unit deformation and $P_{r}$ and $P_{z}$ are the dimensional filling factors. Combining Eq.~\eqref{eq:strain} and ~\eqref{eq:dfdz} and noting that for the WGH$_{16,0,0}$ mode $P_{\epsilon_{r}}\approx P_{z}\approx 0$ and $P_{\epsilon_{z}}\approx P_{r}\approx1$ gives
\begin{equation}
\partial f / \partial P = Q \left(\nu - 0.5 K_{\epsilon_{z}}\right)P_{\epsilon_{z}} / (2\pi V Y),
\label{eq:dfdpcalc}
\end{equation}
which is the radiation pressure induced power to frequency conversion of a WGH mode. Given that~\cite{locke2004} $K_{\epsilon_{z}} = 4.2$, we calculate a radiation pressure induced power to frequency sensitivity of -0.2~Hz/mW for the SLC under test. This process can be repeated to derive the sensitivity of a WGE mode which was used in previous CSO designs, where the radial electric field is dominant. We obtain the same expression as found by Chang et. al.~\cite{chang1997} only without the factor $a$ which we suspect was included due to a misinterpretation of earlier work by Braginsky~et~al.~\cite{braginsky1985}. Due to the anisotropic strain/frequency behavior of sapphire the WGE modes are inherently more sensitive to radiation pressure effects by a factor of approximately 1.4 compared to WGH modes, assuming that the effective mode volumes, quality factors and appropriate filling factors remain constant.
\begin{figure}[t]
\centering
\includegraphics[scale=0.54,keepaspectratio=true]{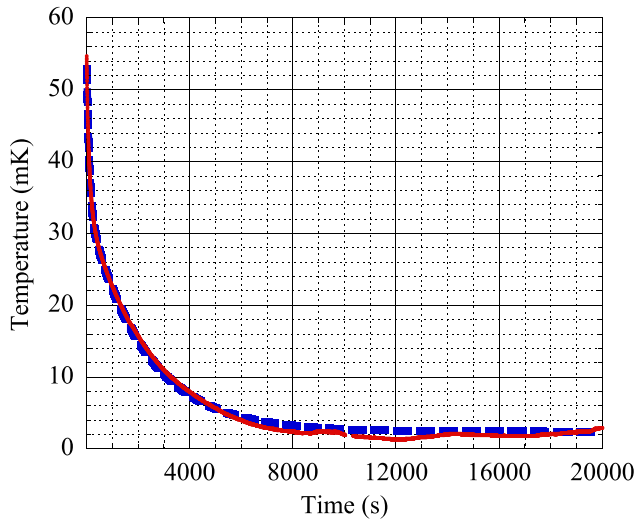}
\caption{(color online) Temperature of the copper post sensor (red data points) as a function of time in response to a 0.4~mW drop in power incident on the resonator. Blue dashed line is Eq.~\eqref{eq:temp} with the values discussed in the text. A linear offset of 5.9~K has been removed.}
\label{fig:temp}
\end{figure}

The power to frequency sensitivity, $\partial f$/$\partial P$, of a WGH mode SLC was measured using the arrangement shown in Fig.~\ref{fig3}, which is the schematic of a typical CSO with the components that would normally form the power control system being used instead to perform this experiment. An arbitrary waveform generator (DDS) supplies a sine wave of fixed amplitude to the bias of a Voltage Controlled Attenuator (VCA), creating a power modulation, $\partial P$, incident on the resonator. This induces frequency variations at the output of the loop oscillator circuit that are of a larger magnitude than the intrinsic fluctuations of both the conventional CSO and a synthesizer externally referenced to a 10~MHz H-maser. The frequency of the beat note between the CSO and the synthesizer is recorded via a frequency counter along with the power incident on an amplitude detector (AD1) prior to entering the SLC, the temperature at the copper post of the cavity (see Fig.~\ref{fig:SLC}) and the heater output of the temperature control system. The readout mixer and frequency counter are immune to amplitude modulation of the CSO output. By computing the Allan deviation of the fractional frequency stability of the beat, we can infer the frequency shift induced by the sine wave modulated power by looking at the beat frequency stability at the integration time corresponding to the period of modulation. Using this method we can measure frequency shifts for modulation frequencies up to 0.1~Hz, which is an upper limit imposed by the frequency counter and data acquisition system - the bandwidth of the Pound system would allow for measurements of modulation frequencies up to tens of Hz. One possible drawback of this approach is that the VCA will induce residual frequency fluctuations as it does not generate a ``pure'' amplitude modulated signal, however for the low modulation frequencies used in this experiment the gain of the Pound frequency control system is sufficiently high enough to suppress this unwanted source of frequency modulation. Performing measurements with the SLC temperature control on and off enables the calculation of independent values of $\partial f$/$\partial P$ due to radiation pressure effects and temperature effects.

\begin{figure}[t]
\centering
\includegraphics[scale=0.54,keepaspectratio=true]{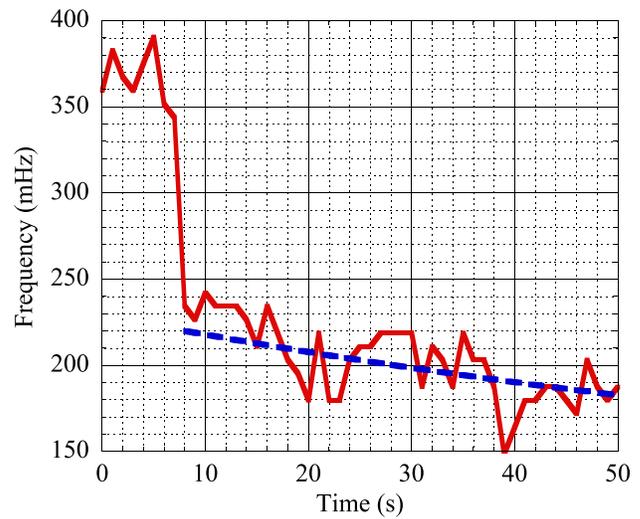}
\caption{(color online) Beat frequency output (red trace) from Fig.~\ref{fig3} as a function of time. Power incident on the resonator increases by 0.4~mW between 7 and 8 seconds. Blue dashed line is Eq.~\eqref{eq:freqmodel}.}
\label{fig:freqshift}
\end{figure}

Initial data was taken with the SLC temperature control deactivated in order to characterize the power to temperature conversion of the resonator. A 20~$\mu$Hz square wave modulation was applied to the VCA bias, inducing a change in incident power of $\delta P$. The temperature response from the sensor in the copper post of the cavity (see Fig.~\ref{fig:SLC}) was observed (Fig.~\ref{fig:temp}), indicating a thermal resistance (R$_{\text{th}}$) between the sapphire and the 4.2~K coldfinger of 134.73$\pm$0.35 K/W. The temperature change can be represented as
\begin{equation}
T\left(t\right) = T\left(t_{0}\right) + \delta P \text{R}_{\text{th}}\left(a_{1} e^{-\frac{t}{\tau_{1}}}+a_{2} e^{-\frac{t}{\tau_{2}}}-1\right),
\label{eq:temp}
\end{equation}
where $a_{1}$ and $a_{2}$ are material weighting factors, such that $a_{1}+a_{2}=1$. For this cavity, based on the data in Fig.~\ref{fig:temp}, the dominant contributions come from the stainless steel coldfinger washer ($a_{1}=0.6$, $\tau_{1}=2200$~s) and the copper cavity ($a_{2}=0.4$, $\tau_{2}=100$~s). Considering the small volume and large surface area of the silver coating and noting that the specific heats of copper and silver are similar at 5~K~\cite{nist2000}, we ignore the thermal contribution of silver. The time constant of the sapphire is already known~\cite{tobar1994} and will be less than 1~ms. For a SLC resonator operating at a small temperature offset, $\Delta T_{\text{off}}$, in the vicinity of the frequency/temperature turning point the temperature to frequency conversion efficiency, $\partial f/\partial T$, can be approximated~\cite{locke2008} as $\Delta T_{\text{off}} f_{0} 10^{-9}$, where $f_{0}$ is the SLC resonance frequency. Combining this with eq.~\eqref{eq:temp} gives the time dependent $\partial f$/$\partial P$ due to temperature effects,
\begin{equation}
\partial f / \partial P= f_{0} \Delta T_{\text{off}} 10^{-9}\text{R}_{\text{th}}\left(a_{1} e^{-\frac{t}{\tau_{1}}}+a_{2} e^{-\frac{t}{\tau_{2}}}-1\right).
\label{eq:freqmodel}
\end{equation}

Figure~\ref{fig:freqshift} shows how the beat frequency responds to stepwise change in the power incident on the resonator. There is a corresponding instantaneous frequency shift due to the change in radiation pressure followed by a gradual change in frequency as the temperature of the resonator changes. The blue dashed line is generated using Eq.~\eqref{eq:freqmodel}.

\begin{figure}[t!]
\centering
\includegraphics[scale=0.54,keepaspectratio=true]{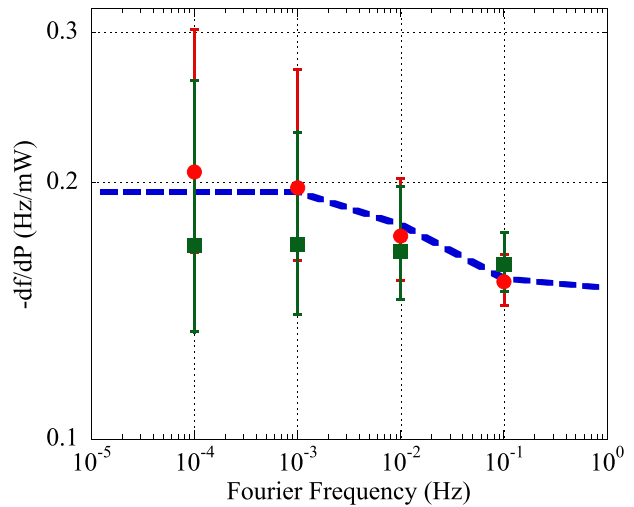}
\caption{(color online) $\partial f/\partial P$ as a function of Fourier frequency obtained using the method shown in Fig.~\ref{fig3}. Data was taken at the frequency/temperature turning point with temperature control system off (red circles) and on (green squares). The blue dashed line is calculated using Eq.~\eqref{eq:freqmodel}.}
\label{fig:dfdp}
\end{figure}

Values of $\partial f/\partial P$ as a function of Fourier frequency are displayed in Fig.~\ref{fig:dfdp}. These were obtained via the method discussed earlier (Fig.~\ref{fig3}), where the power incident on the resonator is modulated by a sine wave and the resulting frequency shift is calculated from the Allan deviation of the fractional frequency stability at the appropriate integration time. As the power modulation frequency decreases the temperature control system becomes increasingly effective, data collected with active temperature control (green squares) gives a value of $\partial f/\partial P$ that is independent of Fourier frequency while values calculated from data without any temperature control (red circles) show an increase in $\partial f/\partial P$ as predicted by Eq.~\eqref{eq:freqmodel} (blue dashed line in Fig.~\ref{fig:dfdp}). Given the 100~s time constant associated with the copper cavity the temperature control system is most effective at Fourier frequencies below 10~mHz. Measurements were also made with temperature control enabled at a setpoint up to 2~K away from the frequency/temperature turning point with no significant change in $\partial f / \partial P$ observed. The error bars presented in Fig.~\ref{fig:dfdp} do not account for uncertainty in estimating the losses in the loop oscillator circuit from prior to the amplitude detector (AD1 in Fig.~\ref{fig3}) to the SLC. We have attributed a power loss of 0.5~dB for the 10~dB directional coupler, 1~dB for losses in the co-axial cables, 0.5~dB for losses in the vacuum can feedthrough and 0.5~dB for losses in the circulator. If the path was lossless then the values presented in Fig.~\ref{fig:dfdp} would increase by a factor of 1.4. Considering that the WGH mode of the SLC used in this work is less sensitive to $\partial f$/$\partial P$ than WGE modes the values obtained are consistent with previous measurements~\cite{chang1997}. 

We also attempted a second measurement technique that used a Fast Fourier Transform vector signal analyzer (FFT) to provide white noise (from 0.1 to 1~Hz) to the VCA bias and then locked the synthesizer frequency to the beat output with a Phase Locked Loop (PLL) filter and then used the FFT to compute the complex transfer function from the filter correction voltage (which is proportional to the SLC resonance frequency) to the VCA bias (which is proportional to the power incident on the SLC). From this one can infer $\partial f$/$\partial P$ for Fourier frequencies from 0.1 to 1~Hz, where the temperature control system has no effect. We confirm that $\partial f$/$\partial P$ due to the radiation pressure effect is independent of both temperature and modulation frequency. Unfortunately this method is more susceptible to sources of systematic error (due to PLL sensitivity to power variations of the CSO output) and thus likely to overestimate the magnitude of $\partial f$/$\partial P$, as such the data has not been presented in Fig.~\ref{fig:dfdp}.

Given that the sign of $\partial f/\partial P$ caused by changes in the magnetic susceptibility of sapphire is opposite to the sign of $\partial f/\partial P$ due to changes in radiation pressure there will exist some temperature below the frequency/temperature turning point where these two effects will negate each other over the appropriate time scales. For the SLC studied in this paper, this temperature is approximately 0.1~K below the frequency/temperature turning point to allow for cancellation at a Fourier frequency of $10^{-3}$ Hz. It could be possible to operate the CSO at this temperature without active temperature control or power control and still achieve reasonable fractional frequency stability at integration times corresponding to the suppressed Fourier frequency. 

In terms of CSO performance the fractional frequency stability limit due to power to frequency conversion will ultimately be dictated by the quality of the power control system. This white noise floor imposed by the power control system can be calculated from
\begin{equation}
\sigma_{y}=2\sqrt{2 ln 2} \sqrt{\alpha} (P_{inc}/ f_{0}) \left|\partial f / \partial P\right|,
\end{equation}
where $\alpha$ characterizes the flicker noise of the power control system amplitude detector and is found by fitting a power law of the form $\alpha/\sqrt{F}$ to the power spectral density of the intrinsic voltage noise floor of the detector. The amplitude detectors currently deployed in the CSO power control system utilize tunnel diodes from Herotek, Inc. (model DT8016). At cryogenic temperatures $\alpha\approx10^{-9.6}$, which assuming $P_{inc}\approx0.2$~mW and $\partial f/\partial P=-0.15$~Hz/mW gives a fractional frequency stability of $\approx1\times10^{-16}$. This is the current limit of CSO performance~\cite{hartnitin32012}, as such any further improvements will require advances to the power control system. Preliminary work shows that a room temperature microwave interferometer amplitude detection circuit~\cite{sann1968,ivanov1998} that produces a voltage output proportional to the amplitude fluctuations of an incident signal exhibits an $\alpha$ of $10^{-12.3}$ which translates into a fractional frequency stability limit of $5\times10^{-18}$.

The authors thank D.L. Creedon for assistance with preparing Fig.~\ref{fig:SLC}. This work was funded by Australian Research Council grants LP0883292, DP130100205, CE110001013 and FL0992016 with support from UWA.


\begin{thebibliography}{99}

\bibitem{hartnett2006} J.G. Hartnett, C.R. Locke, E.N. Ivanov, M.E. Tobar, and P.L. Stanwix, \textit{Appl. Phys. Lett.}, \textbf{89}, 203513 (2006).

\bibitem{locke2008} C.R. Locke, E.N. Ivanov, J.G. Hartnett, P.L. Stanwix, and M.E. Tobar, \textit{Rev. Sci. Instrum.}, \textbf{79}(5), 051301 (2008).

\bibitem{hartnitin32012} J.G. Hartnett, N.R. Nand, and C. Lu, \textit{Appl. Phys. Lett.}, \textbf{100}, 183501 (2012).

\bibitem{pound1946} R.V. Pound, \textit{Rev. Sci. Instrum.}, \textbf{17}, pp. 490--505 (1946).

\bibitem{fortier2011} T.M. Fortier, M.S. Kirchner, F. Quinlan, J. Taylor, J.C. Bergquist, T.Rosenband, N. Lemke, A. Ludlow, Y. Jiang, C.W. Oates, and S.A. Diddams, \textit{Nature Photonics}, \textbf{5}, 425-429 (2011).

\bibitem{wang2010} C. Wang and J.G. Hartnett, \textit{Cryogenics}, \textbf{50}, 336--341 (2010).

\bibitem{hartnett1999} J.G. Hartnett, M.E. Tobar, A.G. Mann, E.N. Ivanov, J. Krupka, and R. Geyer, \textit{IEEE Trans. Ultrason. Ferroelectr. Freq. Control}, \textbf{46}(4), 993-1000 (1999),

\bibitem{braginsky1985} V.B. Braginsky, V.P. Mitrafanov, and V.I. Panov, \textit{Systems with Small Dissipation}, University of Chicago, Chicago, (1985).

\bibitem{chang1997} S. Chang, A.G. Mann, A.N. Luiten, and D.G. Blair, \textit{Phy. Rev. Lett.}, \textbf{79}(11), 2141--2144 (1997).

\bibitem{locke2004} C.R. Locke and M.E. Tobar, \textit{Measurement Science and Technology}, \textbf{15}(10), 2145 (2004).

\bibitem{nist2000} E.D. Marquardt, J.P. Le, and R. Radebaugh, \textit{NIST}, (2000).

\bibitem{tobar1994} M.E. Tobar, E.N. Ivanov, R.A. Woode, and J.H. Searls, \textit{Proceedings of the 1994 IEEE International Frequency Control Symposium} (1994), 433-440.

\bibitem{sann1968} K.H. Sann, \textit{IEEE Trans. Microwave Theory Tech.}, \textbf{MTT-16}, 761--765 (1968).

\bibitem{ivanov1998} E.N. Ivanov, M.E. Tobar, \textit{IEEE Trans. Ultrason. Ferroelectr. Freq. Control}, \textbf{45}, 1526--1536 (1998).

\end{thebibliography}
\end{document}